\documentclass[]{aastex}
\usepackage{spr-astr-addons}

\usepackage{url}
\RequirePackage{color}

\usepackage{longtable}
\usepackage{natbib}
\usepackage{amsmath}
\usepackage{txfonts}
\usepackage{graphicx}
\usepackage{amssymb}
\usepackage[labelfont=bf]{caption}
\begin{document}

\title{Stellar populations of classical and pseudo-bulges for a sample of isolated spiral galaxies}
\shorttitle{Stellar populations of bulges in isolated spiral galaxies}
\shortauthors{Y. Zhao}

\author{Yinghe Zhao\altaffilmark{1, 2}}
\affil{$^1$Purple Mountain Observatory, Chinese Academy of Sciences, Nanjing 210008, China}
\affil{$^2$Key Laboratory of Radio Astronomy , Chinese Academy of Sciences, Nanjing 210008, China}
\altaffiltext{}{e-mail: yhzhao@pmo.ac.cn}

\begin{abstract}
 In this paper we present the stellar population synthesis results for a sample of 75 bulges in isolated spiral Sb-Sc galaxies, using the spectroscopic data from the Sloan Digital Sky Survey and the STARLIGHT code. We find that both pseudo-bulges and classical bulges in our sample are predominantly composed of old stellar populations, with mean mass-weighted stellar age around 10 Gyr. While the stellar population of pseudo-bulges is, in general, younger than that of classical bulges, the difference is not significant, which indicates that it is hard to distinguish pseudo-bulges from classical bulges, at least for these isolated galaxies, only based on their stellar populations. Pseudo-bulges have star formation activities with relatively longer timescale than classical bulges, indicating that secular evolution is more important in this kind of systems. Our results also show that pseudo-bulges  have a lower stellar velocity dispersion than their classical counterparts, which suggests that classical bulges are more dispersion-supported than pseudo-bulges. 

\end{abstract}
\keywords{galaxies: spiral--galaxies: evolution--galaxies: stellar content--galaxies: bulges}
\section{Introduction}
\label{sect:intro}

The properties of bulges, such as their structure, kinematics, and stellar population, are important to probe the physical mechanisms responsible for the formation and evolution of galaxies. Similarities between the global properties of many bulges and of elliptical galaxies have long been recognized \citep[e.g.][]{kor85,ben92}. However, recent observations have revealed that some bulges are more complicated than previously thought and may be formed from spiral disks \citep[e.g.][]{fis06}. In the literature, bulges those appear very similar to pure elliptical systems are named as classical bulges and those relate to the disk are called as pseudo-bulges \citep[e.g.][]{kor04}.

Classical bulges are typically having hot stellar dynamics and more nearly de Vaucauleurs $R^{1/4}$ surface brightness profiles \citep{kor04}. They have nearly the same fundamental plane relation as ellipticals \citep{ben92,ben93}. Pseudo-bulges are flat components with nearly exponential surface brightness profiles and thus more disk-like in both their morphology and shape \citep{fis08}, and they are dominated by rotation in dynamics \citep{kor93,kor04}. However, there remain many uncertainties in making a clear-cut distinction between these two cases, particularly in regard to the stellar populations of spiral bulges.

In the current paradigm, formation scenarios for bulges can be divided into two categories: one is identical to those for pure ellipticals and the other is to involve the secular evolution  \citep[see][ for a review]{kor04}. Classical bulges were formed through rapid and/or violent process which includes both the monolithic collapse of a primordial gas cloud \citep[e.g.][]{lar74} and major/minor merging events \citep{kau96}. While in the secular evolution scenario, bulges have been slowly assembled by internal and environmental secular processes. Stellar population studies can potentially discern between different formation mechanisms responsible for spiral bulges. The detailed analysis of the stellar populations of nearby galaxies can be used to probe their dominant mechanism(s) at the epochs of star formation and mass assembly. Moreover, a successful formation scenario has to reproduce the observed properties of ages, metallicities, and kinematics of the bulges. The light- and mass-weighted quantities can be used to form a comprehensive picture of the star formation history (SFH) of a given system.

In this work, we present a spectroscopic study of the bulge dominated region of a sample of spiral galaxies selected from a well-defined and representative sample of the most isolated galaxies in the local Universe. Our aim is to estimate the age and metallicity of the stellar population for pseudo-bulges and classical bulges, and therefore to try to disentangle between late slow growth and early rapid assembly of the stellar mass in these two types of bulges. Our study might also shed some light on the effect of environment on bulge formation and evolution. 

This paper is organized as following: In section 2, we describe the galaxy sample and the method for stellar populations. Section 3 presents the detailed results of the stellar population synthesis for pseudo-bulges and classical bulges. In the last section we give a brief discussion and summary for our work. Throughout the paper, we use $H_0=75$ km s$^{-1}$ Mpc$^{-1}$.

\section{Sample and Data Reduction}
\subsection{The Sample}
\begin{figure*}[t]
\centering
{\includegraphics[width=0.75\textwidth,bb=8 75 500 276]{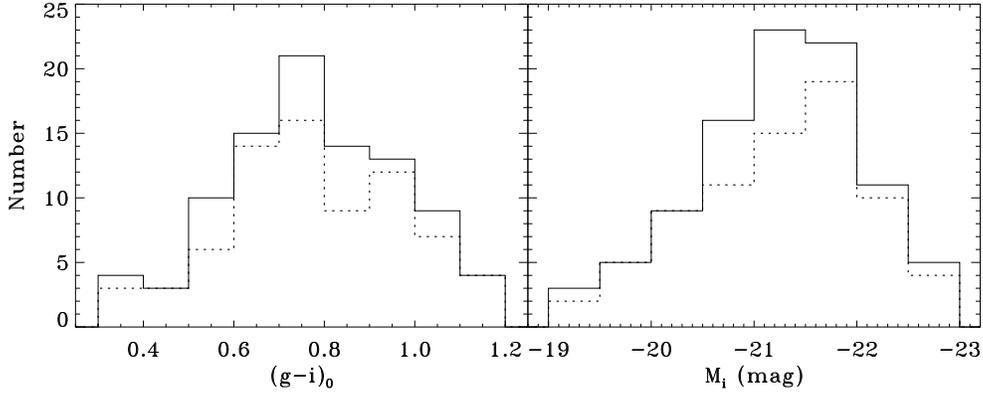}}
\caption{ Number distribution of the isolated spiral galaxies in $(g-i)_0$ color (left panel) and absolute $i$ magnitude (right panel). Solid histogram: DSBV08 sample; dashed histogram: the subsample used in the current work.}
\label{fig1}
\end{figure*}

Based on the Catalog of Isolated Galaxies \citep{kar73} and reevaluated morphologically in the context of the Analysis of the interstellar Medium of Isolated GAlaxies project, \cite[DSBV08 hereafter]{dur08} selected a representative sample of isolated spiral galaxies to analyze their structural properties. The definition of isolation requires that, for a galaxy of diameter $D$, there is no companion/neighbor with diameter $d$ of $D/4<d<4D$ within a distance of $20D$. The final sample in DSBV08 contains 101 galaxies of morphological types Sb-Sc, and was selected according to the following constraints: (1) $1500 < V_{\rm R} < 10000$ km s$^{-1}$, which could avoid inclusion of local supercluster galaxies and ensure an adequate resolution of the image from the Sloan Digital Sky Survey (SDSS), (2) blue-corrected magnitudes (Verdes-Montenegro et al. 2005) $m_{{\rm Bcorr}} < 15$, (3) inclination $<70^\circ$, and (4) available images in SDSS Data Release 6 \citep[Dr6;][]{ade08}.

The galaxy sample used in the present work is the result of cross-correlating the DSBV08's sample with the spectroscopic data of the SDSS DR7 \citep{aba09}, and includes 75 member galaxies. In order to check whether this subsample is representative of the total sample of DSBV08 with respect to the extinction-corrected, integrated $(g-i)_0$  color and the absolute $i$-band magnitude ($M_i$), we plot the $(g-i)_0$ and $M_i$ distributions for the DSBV08 sample and our sample using solid and dashed histograms, respectively, in Figure 1. We can see that ours has similar distributions of $(g-i)_0$ and $M_i$ to the DSBV08 sample, except that our sample contains a smaller fraction of galaxies with $-21.5 <M_i < \sim -20.5$ mag. The $g$- and $i$-band photometric data, as well as the structural parameters (such as the effective surface brightness $\mu_{\rm e}$, the S\'ersic index $n_{\rm b}$ and the effective radius $r_{\rm e}$ (for the bulge), and the central surface brightness $\mu_0$ and the scale length $h_{\rm d}$ (for the disk)) used in the following parts of this paper, are all adopted from DSBV08.

\subsection{Stellar Population Synthesis}
\begin{figure*}[tb]
\centering
\includegraphics[width=0.75\textwidth,bb=20 237 568 625]{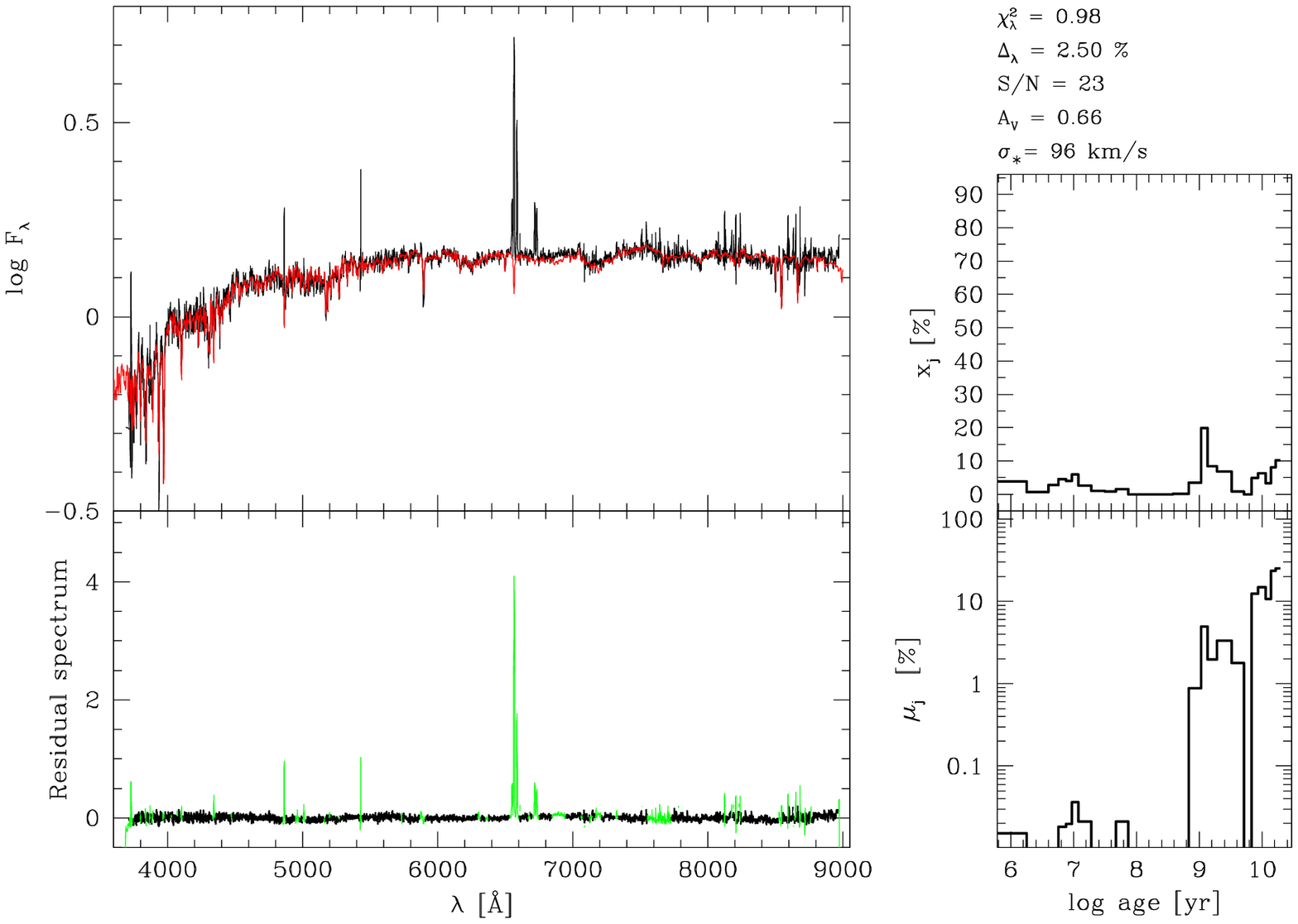}
\caption{Results of the spectral fitting for KIG287. The top left panel shows the logarithm of the observed ($F_\lambda^O$; black line) and the synthetic ($F_\lambda^M$; red line) spectra. The $F_\lambda^O - F_\lambda^M$ residual spectrum is shown in the bottom left panel. Spectral regions actually used in the synthesis are plotted with a black line, while masked regions are plotted with a green line. Panels in the right show the population vector binned in the 25 ages of SSPs used in the model library. The top right panel corresponds to the population vector in flux fraction, normalized to $\lambda_0 = 4020$ \AA, while the corresponding mass fractions vector is shown in the bottom right panel.}
\label{Fig2}
\end{figure*}

In order to obtain the stellar populations for the bulges of these spiral galaxies, we here model the stellar contribution in the SDSS spectra through the modified version of the stellar population synthesis code, STARLIGHT\footnote{STARLIGHT \& SEAGal: http://www.starlight.ufsc.br/ } (\citeauthor{cid05} 2005 (Cid05), 2007;  \citeauthor{mat06} \citeyear{mat06}; \citeauthor{asa07} \citeyear{asa07}), which adopted the stellar library from \cite{bru03}. The code does a search for the linear combination of SSPs to match a given observed spectrum ($O_\lambda$). The model spectrum ($M_\lambda$) is:
\begin{center}
$M_\lambda =M_{\lambda_0}
   \left[
   \sum_{j=1}^{N_\star} x_j b_{j,\lambda} r_\lambda
   \right]
   \otimes G(v_\star,\sigma_\star)$,
 \end{center}
where $b_{j,\lambda} \equiv L_\lambda^{SSP}(t_j,Z_j) /L_{\lambda_0}^{SSP}(t_j,Z_j)$ is the spectrum of the $j^{\rm th}$ SSP normalized at $\lambda_0$, $r_\lambda \equiv 10^{-0.4(A_\lambda - A_{\lambda_0})}$ is the reddening term, {\boldmath $x$} is the population vector, $M_{\lambda_0}$ is the synthetic flux at the normalization wavelength, $N_\star$ is the total number of SSPs, $G(v_\star,\sigma_\star)$ is the line-of-sight stellar velocity distribution, modeled as a Gaussian centered at velocity $v_\star$ and broadened by $\sigma_\star$. The match between model and observed spectra is calculated by
\begin{center}
$\chi^2(x,M_{\lambda_0},A_V,v_\star,\sigma_\star) =
   \sum_{\lambda=1}^{N_\lambda}
   \left[
   \left(O_\lambda - M_\lambda \right) w_\lambda
   \right]^2$,
\end{center}
where the weight spectrum $w_\lambda$ is defined as the inverse of the noise in $O_\lambda$. For more details, please refer to Cid05 and \cite{mat06}. The SSP library follows the work of SEAGal Group, and is made up of $N_\star=150$, including 25 ages (from 1 Myr to 18 Gyr) and 6 metallicities ($Z=0.005, 0.02, 0.2, 0.4, 1\, {\rm and}\, 2.5 Z_\odot$). The spectra were computed with the Salpeter (1955) initial mass function (IMF), Padova 1994 models and the STELIB library \citep{leb03}. The intrinsic reddening is modeled by the foreground dust model, using the extinction law of \cite{car89} with $R_V = 3.1$.

The SDSS spectra cover 3800-9200 \AA, with a resolution ($\lambda/\Delta\lambda$) of $1800 < R < 2100$ and sampling of 2.4 pixels per resolution element. The fiber used in the SDSS spectroscopic observations has a diameter of 3$''$ on the sky. Prior to the synthesis, the Galactic extinction has been corrected with a combination of the extinction law of \cite{car89} and the $A_B$ value from \cite{sch98} as listed in NED\footnote{http://nedwww.ipac.caltech.edu/}. The spectra are transformed into the rest frame using the redshifts given in the FITS header. The SSPs are normalized at $\lambda_0 = 4020$ \AA, while the observed spectra are normalized to the median flux between 4010 and 4060 \AA. The signal-to-noise ratio (S/N) is measured in the relatively clean window between 4730 and 4780 \AA. Masks of $20-30$ \AA\ around obvious emission lines are constructed for each object individually, and more weights are given to the strongest stellar absorption features such as Ca\,{\sc ii} K $\lambda$ 3934, and the Ca\,{\sc ii} triplets, that are less affected by nearby emission lines. For our sample, the S/N varies between 12.0 and 60.5, and the median value is 23.5, with more than eighty-five percent $> 20$. Generally, the fitting results for high S/N objects are better than those for low S/N ones. Inspecting the fitting results, we find that the goodness of fitting ($\chi^2$ value) also somewhat depends on the absorption line equivalent widths (e.g. EW of Ca\,{\sc ii} K). A typical example of our fitting result for KIG287 (UGC04624) is shown in Figure 2.

\section{Results and Analysis}
\subsection{Identification of pseudo-bulges}
For data with high physical spatial resolution, such as images observed by the Hubble Space Telescope, previous work have often used morphological features, such as nuclear bars, nuclear spirals, and/or nuclear rings, to identify  a bulge as a pseudo-bulge \citep{kor04,fis06,fis08}. Whereas for images with a relatively lower physical spatial resolution, such as the SDSS data used here, it is difficult to use such method to identify a pseudo-bulge, and criteria based on the photometric analysis of the surface brightness profile have also been proposed to distinguish pseudo- from classical bulges \citep[e.g.][]{kor04,fis08,fis10,gad09}. \cite{kor04} suggest that pseudo-bulge has a S\'ersic index $n_{\rm b}\simeq1$ to 2. \cite{fis08,fis10} find that more than 90\% of pseudo-bulges have S\'esic index $n_{\rm b}<2$, and all classical bulges have S\'esic index $n_{\rm b}>2$, both in the optical and in the near-infrared, and therefore they propose that the S\'esic index can be used to classify bulges. Hereafter we refer to this method as M01. By comparing the Kormendy relation \citep{kor77}, i.e., the $\left\langle \mu_{\rm e} \right\rangle - r_{\rm e}$ relation (where $\left\langle \mu_{\rm e} \right\rangle$ is the mean effective surface brightness within the effective radius, $r_{\rm e}$), of bulges to that of ellipticals, \cite{gad09} proposed a method (hereafter as ``M02") to distinguish pseudo-bulges from classical bulges, i.e. the pseudo-bulges satisfy the following inequality:
\begin{equation}
\left\langle \mu_{\rm e} \right\rangle > 13.95 + 1.74\times r_{\rm e}
\end{equation}
where $\left\langle \mu_{\rm e} \right\rangle$ and $r_{\rm e}$ are measured in the SDSS $i$-band images, and $r_{\rm e}$ is in units of parsec.

Therefore, we can classify these bulges using the above two methods. However, we need to check whether our data are suitable for identifying pseudo-bulges, i.e., whether the structural parameters for these bulges can be reliably derived. To this purpose, we use the same method as that in \cite{gad08}, and calculate the ratio ($R_{\rm{psf}}$) between bulge effective radius and SDSS PSF half width of half maximum (HWHM). As pointed out in \cite{gad08}, if $r_{\rm e}$ is larger than $\sim$80 percent of the PSF HWHM, the derived structural properties are reliable. In the left panel of Figure 3, we show the distribution of $R_{\rm{psf}}$. We can find that all of the bulges have effective radii above 0.8 times the PSF HWHM, and that only one bulge has its effective radius below 1 times the PSF HWHM. Therefore, it is reasonable to use this sample to distinguish pseudo-bulges from classical bulges.

The right panel of Figure 3 plots $\left\langle \mu_{\rm e} \right\rangle$ (calculated from $\mu_{\rm e}$ and $n_{\rm b}$ using equation (9) in \cite{gra05}) against $r_{\rm e}$ for all bulges in our sample. Bulges with S\'esic index above and below 2 are given by solid and open symbols respectively.  According to method M01, there are 14 classical bulges ($n_{\rm b}>2$) and 61 pseudo-bulges ($n_{\rm b}<2$) in our sample. The solid line in Fig. 3 gives the dividing line between pseudo-bulges and classical bulges based on method M02, which results in 23 classical bulges and 52 pseudo-bulges. We can find that most M02-based pseudo-bulges are consistent with M01-based results, while M02-based classical bulge sample is about 1.6 times large than M01-based one. This difference will affect our results to some extent, and we'll discuss it later.

\begin{figure*}[t]
\centering
{\includegraphics[width=0.75\textwidth,bb=20 76 496 256]{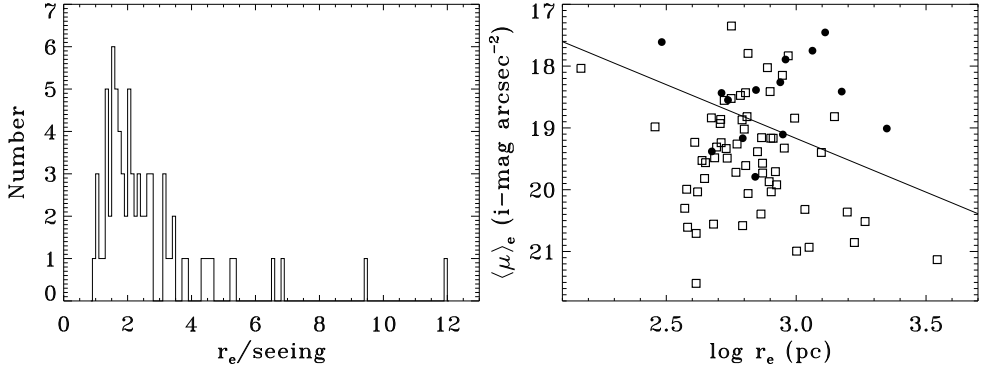}}
\caption{{\it Left panel}: Distribution of the ratio between bulge effective radius and PSF HWHM for all galaxies in our sample. Only one bulge has its effective radius below 1 times the PSF HWHM. {\it Right panel}: The mean effective surface brightness within the effective radius plotted against the logarithm of the effective radius, i.e. the Kormendy relation \citep{kor77}. Bulges with S\'esic index above and below 2 are shown with solid and open symbols, respectively. The solid line is used to separate the bulges into two groups: pseudo (below the line) and classical.}
\label{fig3}
\end{figure*}

\subsection{SFHs from STARLIGHT Fitting}
Based on the fitting results of STARLIGHT, we can obtain/derive the following parameters for the bulges in our sample:  mean stellar age, mean stellar metallicity, the contribution of flux and mass from different SSPs, and stellar velocity dispersion. Both mass- and light-weighted mean stellar ages and metallicities are estimated. These properties can provide us very useful probes for the SFH studies.

\subsubsection{Mean Stellar Age and Metallicity}
In the left panels of Figure 4, we plot the distributions of two mean stellar ages (mass- and light-weighted mean ages; Cid05) estimated for all of the bulges in our sample, as the dashed and solid histograms show the results for pseudo-bulges and classical bulges, respectively. The mass-weighted mean stellar age is defined as,
\begin{equation}
\left<\log\, t_\star \right>_M = \sum_{j=1}^{N_\star} \mu_j\, \log\, t_j,
\end{equation}
and the light-weighted mean stellar age is,
\begin{equation}
\left<\log\, t_\star \right>_L = \sum_{j=1}^{N_\star} x_j\, \log\, t_j.
\end{equation}
where $\mu_j$ and $x_j$ represent the fractional contributions to the stellar mass and luminosity of the SSP with age $t_j$ respectively. $N_\star$ is the number of SSPs. It is easy to understand that $\left< t_\star \right>_M$ is associated with the mass assembly history of galaxies, whereas $\left<t_\star \right>_L$ is strongly affected by the recent SFH of a given galaxy. The fitting results of these two kinds of ages for our bulges are summarized in Table 1. According to Cid05 and  \cite{mat06}, the uncertainties of these two parameters depend on the S/N of the input spectra. In general, the rms of the fitted $\left<\log\, t_\star \right>_M$ is $\sim 0.1$ dex for $\rm{S/N} > 10$, while the rms of the fitted $\left<\log\, t_\star \right>_L$ is $< 0.1$ dex for $\rm{S/N} > 10$.

\begin{table*}[t]
\label{table1}
\begin{center}
\caption[]{Average values of derived parameters for M02-based pseudo-bulges and classical bulges, with corresponding standard deviation given in the parentheses.}
\begin{tabular}{cccccccc}
\hline\noalign{\smallskip}
&& $\left<\log\,t_\star \right>$ & $\left<Z_\star \right>$ & $\left<f_{\rm Y}\right>$ & $\left<f_{\rm M}\right>$ & $\left<f_{\rm O}\right>$ &$\left<\sigma_\star\right>$\\
Bulge$^{a}$&Number &(yr)&($Z_\odot$)&(\%)&(\%)&(\%)&(km s$^{-1}$)\\
\hline\noalign{\smallskip}
Classical&23&8.84(0.80)&0.87(0.38)&30.10(25.62)& 7.62(12.05)& 62.28(28.75)&116.7(34.8)\\[-0.1cm]
--&--&10.02(0.09)&1.00(0.25)&0.45(0.78)&1.55(3.20)& 98.00(3.29)&--\\
--&13&8.74(0.91)&0.85(0.36)&34.20(28.60)&6.73(6.70)&59.07(30.35)&129.1(37.1)\\[-0.1cm]
--&--&10.03(0.09)&0.97(0.19)&0.54(0.84)&1.21(1.43)&98.25(1.62)&--\\
\hline\noalign{\smallskip}
Pseudo&52&8.68(0.52)&0.70(0.25)& 34.18(17.89)& 7.93(9.45)&57.89(22.20)&72.4(16.2)\\[-0.1cm]
--&--&9.92(0.13)&0.81(0.29)&0.53(0.77)&2.58(4.15)&96.89(4.77)&--\\
--&7&8.52(0.72)&0.53(0.15)&44.14(22.32)&3.62(3.59)&52.24(24.07)&63.7(18.4)\\[-0.1cm]
--&--&9.89(0.19)&0.68(0.31)&0.90(1.14)&1.83(3.07)&97.27(4.16)&--\\
\hline\noalign{\smallskip}
\end{tabular}
\end{center}
\tablenotetext{a}{Except for $\sigma_\star$, the first and second rows of each type of bulges are the light(-weighted) and mass(-weighted) results, respectively. The two subsamples with small numbers of member galaxies are used to check the aperture effects, see Section 4 for details.}
 \end{table*}
 
\begin{figure*}[t]
\centering
{\includegraphics[width=0.75\textwidth,bb=5 2 500 362]{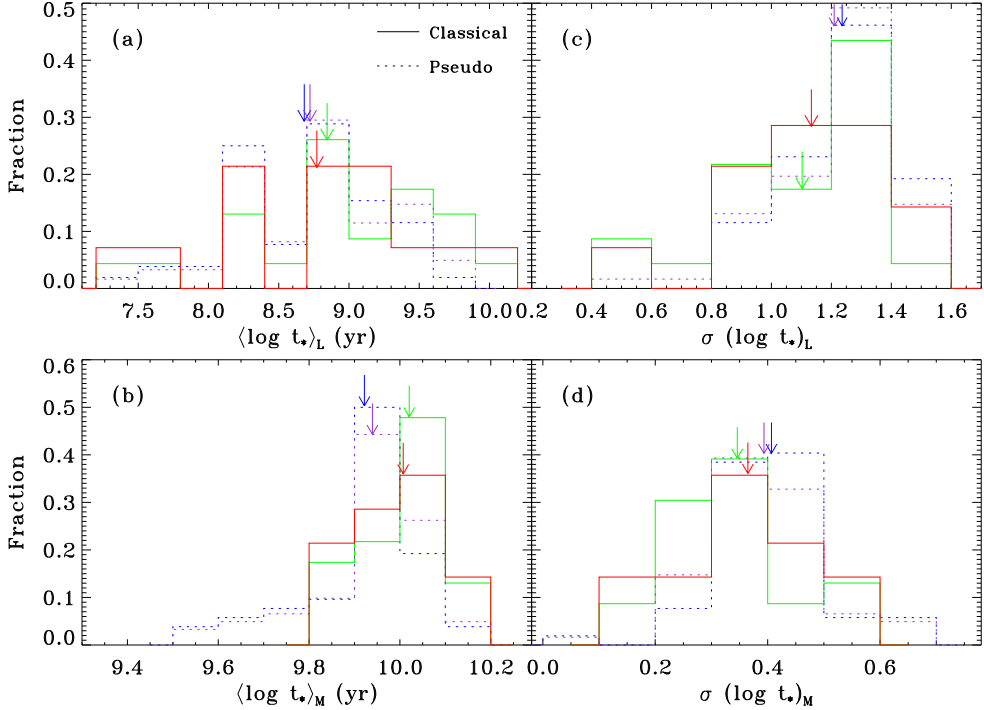}}
\caption{Statistics on the fitting results. In each panel, purple and red histograms are M01-based results, and blue and green histograms are M02-based results, and arrows give the average values for classical (solid-headed arrow) and pseudo bulges. {\it Left panels}: Fraction distributions of the bulges in mean stellar ages. {\it Right panels}: Fraction distributions of the bulges in the dispersions of $\left<\log t_\star\right>$.}
\label{Fig4}
\end{figure*}

\begin{figure*}[t]
\centering
{\includegraphics[width=0.75\textwidth,bb=8 75 500 275]{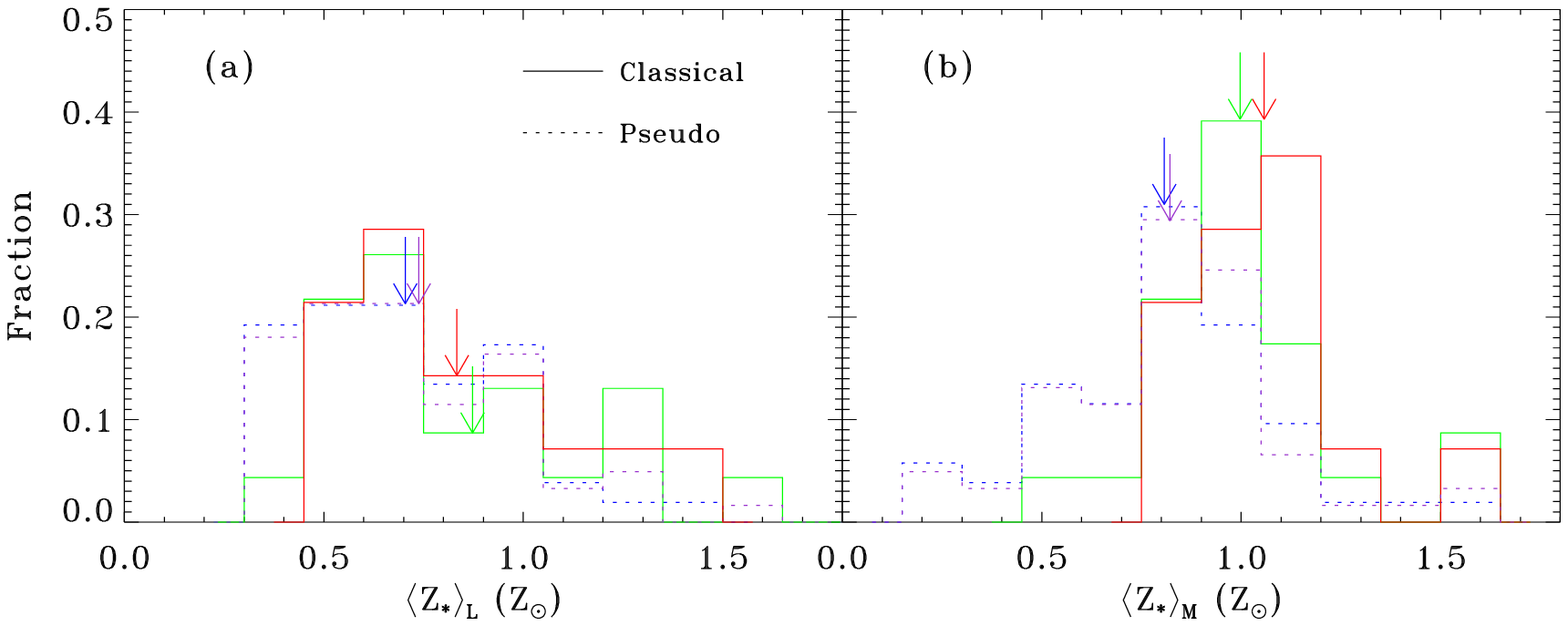}}
\caption{Fraction distributions of the bulges in light- (left panel) and mass-weighted (right panel) mean stellar metallicities. Arrows show the average values for classical (solid-headed arrow) and pseudo bulges. Purple and red histograms are M01-based results, and blue and green histograms are M02-based results.}
\label{fig5}
\end{figure*}

Figs. 4(a) and (b) respectively show the fraction distributions of $\left<\log t_\star \right>_L$ and $\left<\log t_\star \right>_M$, with purple and red lines for M01-based results, and blue and green lines for M02-based results. It is interesting to note that the differences between M01- and M02-based results are very small, and therefore we demonstrate results for bulges classified with both methods but only discuss the M02-based results in the following. However, we should also note that M01-based pseudo-bulges and classical bulges are a bit older and younger, respectively, than M02-based ones,which results in a smaller difference in the averaged SFHs between these two types of bulges. This might be due to that all inactive bulges with $n_{\rm b}<2$ are above the Kormendy relation \citep{fis10}.

From Figure 4 we can see that, in general, both $\left<\log t_\star \right>_L$ and $\left<\log t_\star \right>_M$ of pseudo-bulges tend to be younger than classical bulges, which is verified by the average values (as shown by the arrows in Figs. 4(a) and (b)) of these two stellar ages. As shown in Table 1, the mass-weighted stellar ages of pseudo-bulges and classical bulges are both around 10 Gyr, indicating all bulges are predominantly composed of old components. This result is consistent with the finding of \cite{mac09}, who show that more than 80\% of the stellar mass is contributed by old and metal rich stellar populations for all of the eight bulges in their sample, using integrated spectra. However, the distribution of $\left<\log t_\star \right>_M$ for pseudo-bulges has a tail extending towards $\sim$3 Gyr, while classical bulges have a narrow $\left<\log t_\star \right>_M$ distribution around 10 Gyr, which suggests that pseudo-bulges should have relatively longer mass assembly histories than classical bulges, and secular contributions to the evolution of pseudo-bulges are more important.

To investigate the SFH of bulges in more detail, we calculated the light-weighted and mass-weighted standard deviations of the log age, which are defined as the following (Cid05),
\begin{equation}
\sigma _L (\log t_\star ) = \left[ {\sum\limits_{j = 1}^{N_\star } {x_j (\log t_j  - \left\langle {\log t_\star } \right\rangle _L )^2 } } \right]^{1/2}.
\end{equation}
and
\begin{equation}
\sigma _M (\log t_\star ) = \left[ {\sum\limits_{j = 1}^{N_\star } {\mu_j (\log t_j  - \left\langle {\log t_\star } \right\rangle _M )^2 } } \right]^{1/2}.
\end{equation}
These two higher moments of the age distribution could be used to distinguish systems dominated by a single population from those which
had bursty or continuous SFHs.

The right panels in Figure 4 display the fraction frequency histograms of bulges in these two parameters. Both of the average values of $\sigma _M (\log t_\star )$ and $\sigma _L (\log t_\star )$ are significantly larger than zero, indicating that bulges are not dominated by a single population. At the same time, we can see from Figs. 4(c) and (d) that classical bulges have smaller average $\sigma _L (\log t_\star )$ and $\sigma _M (\log t_\star )$ than pseudo-bulges, which indicates that these two types of bulges have different SFHs.

Similar to $\left<\log t_\star \right>_L$ and $\left<\log t_\star \right>_M$, we can define and derive light- and mass-weighted mean stellar metallicities, $\left<Z_\star \right>_L$ and $\left<Z_\star \right>_M$, which are listed in Table 1. In Figure 5 we compare the distributions of $\left<Z_\star \right>_L$ and $\left<Z_\star \right>_M$ for pseudo-bulges and classical bulges. Pseudo-bulges have mass-weighted mean stellar metallicity less than solar abundance (average value of $\sim 0.8$ $Z_\odot$), while classical bulges have (mass-weighted) metallicity similar to the solar abundance (average value of $\sim 1$ $Z_\odot$).

However, this result might not suggest that pseudo-bulges are generally less abundant in metal than classical bulges. This is because that bulges are known to follow a luminosity-metallicity law \citep[e.g.][]{jab96}, and the lower mean metallicity for pseudo-bulges may be simply due to them having a lower mean luminosity in our sample. To check this, we plot the $i$-band absolute magnitude for all bulges ($M_{i,\,{\rm B}}$, calculated from $\mu_{\rm e}$, $n$ and $R_{\rm e}$ using equations given in \citeauthor{gra05} \citeyear{gra05}) vs $\left<Z_\star \right>_M$ in left panel of Fig. 6, with triangles and circles showing pseudo- and classical bulges respectively. From the figure we can find a weak trend in metallicity with luminosity, with the Spearman rank correlation coefficient ($\rho$) of -0.41 at a significance level of $\sim 3\sigma$. An obvious feature of this relation is that it flattens out at the high luminosity end, which has been shown in \cite{Tre04} for $\sim 51000$ star-forming galaxies (but based on gas phase metallicity). Therefore, the apparent lower mean metallicity of pseudo-bulges, comparing with classical ones, might be due to their relatively smaller mean luminosity, as shown in Fig. 6.

\begin{figure*}[t]
\centering
{\includegraphics[width=0.75\textwidth,bb=2 82 502 278]{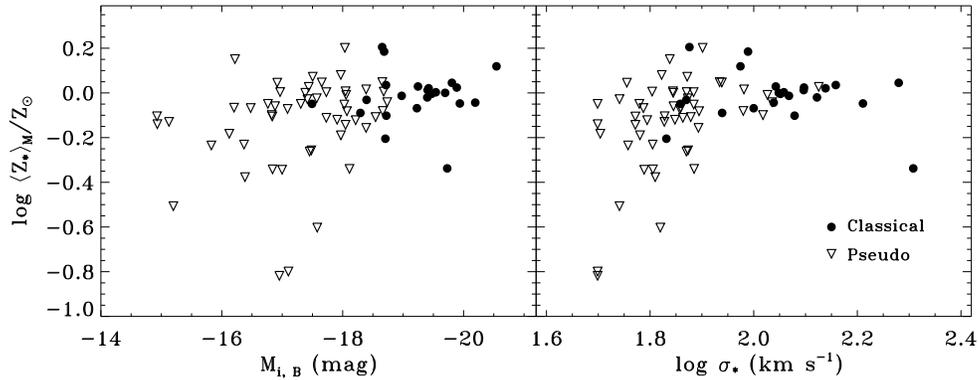}}
\caption{The $i$-band absolute magnitude (left panel) and velocity dispersion (right panel) vs stellar mass-weighted metallicity for pseudo- (open triangle) and classical (solid cirlce) bulges.}
\label{fig6}
\end{figure*} 

\subsubsection{The Approximate SFHs}
STARLIGHT provides us the stellar population vector, for example the fraction of flux contributed by certain SSPs. However, as shown in Cid05, the individual components of {\boldmath $x$} are very uncertain, whereas the binned vectors of {\boldmath $x$}, i.e. `young' ($t_j < 10^8$ yr), `intermediate-age'  ($10^8 \leq t_j \leq 10^9$ yr), and `old' ($t_j > 10^9$ yr) components  ($f_{\rm Y}$, $f_{\rm I}$, and $f_{\rm O}$, respectively), have uncertainties less than 0.05, 0.1 and 0.1, respectively, for S/N$\geq$10. With the fractions of these three stellar populations, a very coarse but robust SFH can be generated and the results are shown in Table 1.

As shown in Table 1, our stellar synthesis result indicates that, in general, classical and pseudo bulges in our sample do not seem to much differ from each other in the stellar populations. Comparing to classical bulges, pseudo-bulges have a bit more contributions from the young (4.1\% for light and 0.1\% for mass) and inter-mediate (0.3\% for light and 1.0\% for mass) components and less contribution from the old component (accordingly, 4.4\% for light and 1.1\% for mass). Our result suggest that it might be hard to distinguish pseudo-bulges from classical bulges only based on their stellar populations.

\subsubsection{Stellar Velocity Dispersion}
\begin{figure*}[t]
\centering
{\includegraphics[width=0.75\textwidth,bb=1 82 505 265]{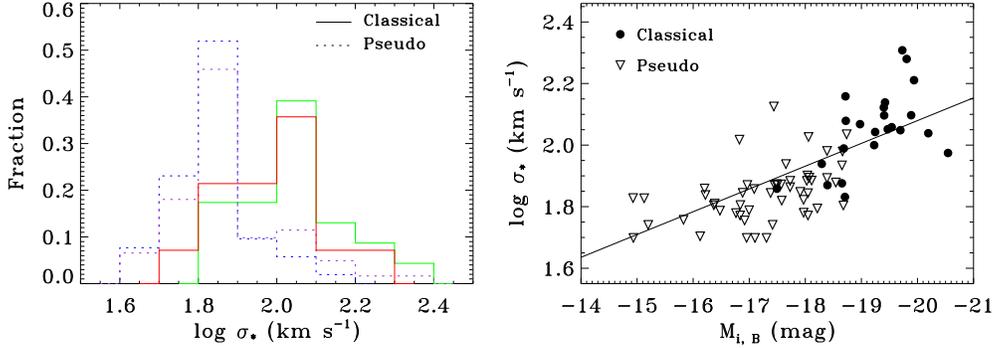}}
\caption{{\em Left panel}: Fraction distribution of the pseudo-bulges and classical bulges in the velocity dispersion. Purple and red histograms are M01-based results, and blue and green histograms are M02-based results. {\em Right panel}: Faber-Jackson relation for all bulges, with the solid line showing the SDSS $i$-band result from \cite{lab10} for early type galaxies.}
\label{fig7}
\end{figure*}
Despite the stellar population vector, STARLIGHT also outputs the broadening parameter, $\sigma_\star$, which depends on the resolution of the SSP library, and the velocity dispersion and instrumental resolution of the input spectra. Due to the limited spectral resolution of the SDSS spectra, it is recommended to use only spectra with signal-to-noise ratio above 10. We have verified that all of the galaxies in our sample comply with this criterion. After correcting for the instrumental resolutions of both the SDSS spectra ($\sigma_{\rm{inst}}\sim 70\,{\rm {km\,s^{-1}}}$) and the STELIB library ($\sim 86\,{\rm {km\,s^{-1}}}$), the derived stellar velocity dispersions for our sample are taken as the central value without further corrections, although these velocity dispersions are obtained through a fixed fiber aperture with diameter of 3$''$. This is because that, for nearby Sa$-$Sd galaxies, observed $\sigma_\star$ profiles are not central peaked but nearly constant in the central 10 arcsec region \citep[e.g.][]{her98,gor07}, and the variation of $\sigma_\star$ measured with different apertures is small \citep{piz04}.

The right panel of Fig. 6 shows the relationship between $\log \left<\sigma_\star\right>$ and $\log \left<Z_\star\right>_M$ for pseudo- and classical bulges. This relation also flattens out at high $\left<\sigma_\star\right>$ end, similar to that between $M_{i,\,{\rm B}}$ and $\log \left<Z_\star\right>_M$. It is not unreasonable because both $M_{i,\,{\rm B}}$ and $\left<\sigma_\star\right>$ can trace the stellar mass for a more fundamental relation, the stellar mass-metallicity relation. The distribution of pseudo- and classical bulges in the $\sigma_\star-Z_\star$ plot further confirms that pseudo- and classical bulges have different mean stellar metallicities may be simply due to they having different stellar masses.

The fraction distributions of the velocity dispersions for pseudo-bulges and classical bulges are plotted in the left panel of Fig. 7. From the figure we can see that about a half of the sampled pseudo-bulges have their $\sigma_\star \lesssim \sigma_{\rm{inst}}$, which will result in a serious uncertainty. However, these measurements are still used as a very rough estimation, and it will not affect our main conclusion that classical bulges have larger velocity dispersions than pseudo-bulges (see Fig. 7) in our sample. As shown in Table 1, the mean values of the velocity dispersions for pseudo ($\left<\sigma_\star\right>_{\rm{p}}$) and classical ($\left<\sigma_\star\right>_{\rm{c}}$) bulges are $72.4\pm16.2\,{\rm{km\,s^{-1}}}$ and $116.6\pm34.8\,{\rm{km\,s^{-1}}}$, respectively. 

One characteristic of pseudo-bulges is their positions with respect to the Faber-Jackson relation \citep[FJ;][]{fab76}, which is a correlation between the central velocity dispersion of elliptical galaxies/bulges and their luminosity. \cite{kor04} show that pseudo-bulges fall well below this relation. In the right panel of Fig. 7, we show the relation between the velocity dispersion and absolute magnitude for all bulges, with the solid line showing the SDSS $i$-band result from \cite{lab10} for early type galaxies. We can see from the figure that, most pseudo-bulges are indeed below the FJ relation, which suggests that they have a lower velocity dispersion comparing to their classical counterparts. Therefore, our result confirms the general finding that classical bulges are more dispersion-supported than pseudo-bulges.

\section{Discussion and Conclusions}
In this paper we present the stellar population synthesis results for a sample of 75 isolated classical bulges and pseudo-bulges, using the SDSS spectra and the STARLIGHT code. For this sample we find that the stellar population of pseudo-bulges is, in general, younger and less abundant in metal than that of classical bulges, while these differences are not significant, and both types of bulges are predominantly composed of old stellar populations, with mean mass-weighted stellar age around 10 Gyr. The apparent lower mean stellar metallicity of pseudo-bulges, comparing with classical ones, may be simply due to that they have relatively smaller mean stellar masses. Pseudo-bulges have star formation activities with relatively longer timescale than classical bulges, indicating that secular evolution is more important in this kind of systems. By comparing the positions of pseudo-bulges with respect to the FJ relation, we confirm the general finding that classical bulges are more dispersion-supported than pseudo-bulges.

However, the spectra for all bulges used in the current work were obtained through a fixed-size aperture, the interpretation of the derived stellar populations is not straightforward as they may be contaminated by the disk population superimposed on the line of sight. Therefore, we need to address the question of how much contamination by the light of the disk can affect our results. In the following, we will discuss the disk contamination.

In the literature, several works \citep[e.g.][]{jab96,pru01} have discussed the estimation of the level of contamination by the disk light. The methods in \cite{jab96} and \cite{pru01} are similar, i.e., the former defined a radius, $R_6$, within which the light from bulge ($L_{\rm b}$) is about 6 times more than that from the disk ($L_{\rm d}$), whereas the latter defined $R'_6$ by $g_{\rm b}(R'_6)/g_{\rm d}(R'_6)=6$, where $g_{\rm b}$ and $g_{\rm d}$ are respectively the growth curves for the bulge and for the disk. The growth curves give the flux integrated in an aperture parameterized by the equivalent radius, i.e. a geometric average of major and minor axes. As pointed out in \cite{pru01}, although the definitions of $R_6$ and $R'_6$ formally differ, in practice the values are close. 

\begin{figure}[t]
\centering
{\includegraphics[width=0.46\textwidth,bb=52 52 408 270]{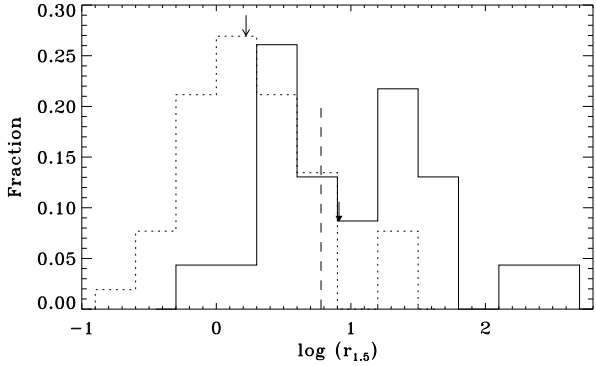}}
\caption{Fraction distributions of the bulge to disk $i$-band luminosity ratio ($r_{1.5}$) for the pseudo-bulges (dashed) and classical bulges (solid), with arrows showing the median values. The vertical dashed line gives the position of $L_{\rm b}/L_{\rm d}=6$.}
\label{fig8}
\end{figure}

In the current work, we use the same idea as Jablonka et al.'s. However, here we calculate the bulge to disk luminosity ratio within $r=1''.5$, i.e. $r_{1.5}=L_{{\rm b}}(1''.5)/L_{{\rm d}}(1''.5)$, instead of $R_6$, for our fixed-size aperture with radius of 1$''$.5. Then we compare $r_{1.5}$ with $L_{\rm b}(R_6)/L_{\rm d}(R_6)=6$. Given the values of $\mu_{\rm e}$, $n_{\rm b}$, $r_{\rm e}, $ $\mu_0$, and $h_{\rm d}$ from DSBV08's photometric decomposition based on the SDSS $i$-band images, $r_{1.5}$ can be calculated using the explicit forms shown in \cite{gra05}. In Figure 8 we demonstrate the fraction distributions of $r_{1.5}$ for pseudo-bulges (dashed line) and classical bulges (solid line), with the superimposed vertical dashed line giving the position of $L_{\rm b}/L_{\rm d}=6$. We can see that the media values of $r_{1.5}$ (arrows in the figure) are 1.7 and 8.2, for pseudo-bulges and classical bulges respectively, which indicates that about a half of the spectra for the pseudo-bulge sample, whereas only a small fraction of the classical bulge sample, are seriously contaminated ($L_{\rm b}/L_{\rm d}<2$) by the light from the disk. Therefore, we need to check to what extent our results can be reliably retrieved.

To the above purpose, we draw subsamples with $r_{1.5} \geq 6$ both for the pseudo-bulge and classical bulge samples. As listed in Table 1, the subsamples of pseudo-bulges and classical bulges contain 7 and 13 member galaxies, respectively. The mean values of the parameters presented in Section 3.2 for these two subsamples are also given in Table 1. It is interesting to find that, in general, the differences of the derived parameters between the two subsamples and their parent samples are small, which indicates that  the bulge and inner disk might have similar stellar populations. Therefore, our results might not be much affected by the contamination by the disk. However, these two subsamples, especially the pseudo-bulge subsample, are much smaller than their parent samples, which can lead to much uncertainty. Our results need to be checked using a much larger sample that covers the entire Hubble types of spiral galaxies, while already these results provide important clues for bulge formation and evolution models.

\acknowledgements
Y. Zhao is grateful for the financial supports from the NSF of China (grant No. 10903029), and greatly appreciates the anonymous referee for her/his careful reading and constructive comments. The \emph{starlight} project is supported by the Brazilian agencies CNPq, CAPES and FAPESP and by the  France-Brazil CAPES/Cofecub program. This research has made use of the NASA/IPAC Extragalactic Database (NED), which is operated by the Jet Propulsion Laboratory, California Institute of Technology, under contract with the National Aeronautics and Space Administration.  All the authors acknowledge the work of the Sloan Digital Sky Survey (SDSS) team. Funding for the SDSS and SDSS-II has been provided by the Alfred P. Sloan Foundation, the Participating Institutions, the National Science Foundation, the U.S. Department of Energy, the National Aeronautics and Space Administration, the Japanese Monbukagakusho, the Max Planck Society, and the Higher Education Funding Council for England. The SDSS Web Site is http://www.sdss.org/. The SDSS is managed by the Astrophysical Research Consortium for the Participating Institutions. The Participating Institutions are the American Museum of Natural History, Astrophysical Institute Potsdam, University of Basel, University of Cambridge, Case Western Reserve University, University of Chicago, Drexel University, Fermilab, the Institute for Advanced Study, the Japan Participation Group, Johns Hopkins University, the Joint Institute for Nuclear Astrophysics, the Kavli Institute for Particle Astrophysics and Cosmology, the Korean Scientist Group, the Chinese Academy of Sciences (LAMOST), Los Alamos National Laboratory, the Max-Planck Institute for Astronomy (MPIA), the Max-Planck-Institute for Astrophysics (MPA), New Mexico State University, Ohio State University, University of Pittsburgh, University of Portsmouth, Princeton University, the United States Naval Observatory, and the University of Washington.

\end{document}